\documentclass[a4j,11pt]{article}

\usepackage{amsthm}
\usepackage{amsmath}
\usepackage{amssymb}
\usepackage{amsfonts}
\usepackage{wrapfig}
\usepackage[pdftex]{graphicx,color}
\usepackage{comment}
\usepackage[hidelinks]{hyperref}
\usepackage{url}

\theoremstyle{plain}

\newtheorem{theorem}{Theorem}

\newtheorem{remark}{Remark}
\newtheorem{corollary}{Corollary}

\newcommand{\argmax}{\mathop{\rm arg~max}\limits}

\begin{document}

\title{Multi-Goal Prior Selection: A Way to Reconcile Bayesian and Classical Approaches for Random Effects Models}

\author{Masayo Y. Hirose \\{The Institute of Statistical Mathematics, Japan}\\
    and \\
    Partha Lahiri\\
    Joint Program in Survey Methodology \& Department of Mathematics,\\
    University of Maryland, College Park, U.S.
 }
\date{\hfill}

\maketitle

\begin{abstract}
The two-level normal hierarchical model has played an important role in statistical theory and applications. In this paper, we first introduce a general adjusted maximum likelihood method for estimating the unknown variance component of the model and the associated empirical best linear unbiased predictor of the random effects.  We then discuss a new idea for selecting prior for the hyperparameters.   The prior, called a multi-goal prior, produces Bayesian solutions for hyperparmeters and random effects that match (in the higher order asymptotic sense) the corresponding classical solution in linear mixed model with respect to several properties.  Moreover, we establish for the first time an analytical equivalence of the posterior variances under the proposed multi-goal prior and the corresponding parametric bootstrap second-order mean squared error estimates in the context of a random effects model.

\end{abstract}


\section{Introduction}

Simultaneous estimation of several independent normal means has been a topic of great research interest, especially in the 60's, 70's and 80's, after the publication of the celebrated James-Stein estimator (James and Stein, 1961). 
Let $y=(y_1,\ldots,y_m)^{\prime}$ be a maximum likelihood estimator of $\theta=(\theta_1,\cdots,\theta_m)^{\prime}$ under the model:
$y_i|\theta_i \stackrel{ind.}{\sim} N(\theta_i, 1),\; i=1,\cdots,m.$
James-Stein (1961) provided a surprising result that for $m\ge 3$, $y$ is an inadmissible estimator of $\theta$ under the model and the sum of squared error loss function: $L(\hat{\theta},\theta)=\sum_{i=1}^{m}(\hat{\theta}_i-\theta_i)^2$. They also showed 
that the estimator $\hat \theta_i^{JS}=(1-\hat{B}^{JS})y_i$, where $\hat{B}^{JS}={(m-2)}/{(\sum_{i=1}^{m}y_i^2)}$, dominates $y$ in terms of the frequentist's risk.  To be specific, 
$E[\sum_i^{m}(\hat{\theta}_i^{JS}-\theta_i)^2|\theta]\leq E[\sum_i^{m}(y_i-\theta_i)^2|\theta]$, for all 
$\theta \in \mathcal{R}^m,$  the $m$-dimensional Euclidean space, with strict inequality holding for at least one point $\theta$. 

The potential of different extensions of the James-Stein estimator to improve data analysis became transparent when Efron and Morris (1973) provided an empirical Bayesian justification of the James-Stein estimator using the prior $\theta_i\sim^{iid.}N(0,A)$, \; $i=1,\cdots,m$. 
Some earlier applications of empirical Bayesian method include the estimation of: (i) false alarm probabilities in New York City (Carter and Rolph, 1974), (ii) the batting averages of major league baseball players (Efron and Morris, 1975), (iii) prevalence of toxoplasmosis in El Salvador (Efron and Morris, 1975) and (iv) per-capita income of small places in the USA (Fay and Herriott, 1979). More recently, variants of the method given in Efron and Morris (1973) was used: to estimate poverty rates for the US states, counties, and school districts (Citro and Kalton, 2000) and Chilean municipalities (Casas-Cordero, Encina and Lahiri , 2016), and to estimate proportions at the
lowest level of literacy for states and counties (Mohadjer et al. 2012).

The following two-level Normal hierarchical model is an extension of the model used by Efron and Morris (1973):

\noindent For $i=1,\ldots, m$, 

{\rm Level \ 1 \ (sampling \ model):}
$y_i|\theta _i \stackrel{\mathrm{ind.}}{\sim} N(\theta_i,D_i)$;

{\rm Level \ 2 \ (linking \ model):}
$\theta_i \stackrel{\mathrm{ind.}}{\sim}N(x_i^{\prime}\beta, A)$.

\noindent In the above model, level 1 is used to account for the sampling distribution of unbiased estimates $y_i$ based on observations taken from the $i$th population. In this model, we assume that the sampling variances $D_i$ are known and this assumption often follows from the asymptotic variances of transformed direct
 estimates (Efron and Morris, 1975; Carter and Rolph, 1974) or from empirical variance modeling (Fay and Herriot, 1979, Otto and Bell, 1995).
Level 2 links the random effects $\theta_i$ to a vector of $p$ known auxiliary variables $x_i=(x_{i1},\cdots,x_{ip})^{\prime}$, which are often obtained from various alternative data sources. The parameters $\beta$ and $A$ are generally unknown and are estimated from the available data. We assume that $\beta\in \mathcal{R}^p,$ the $p$-dimensional Euclidian space. 
In the growing field of small area estimation, this model is commonly referred to as the Fay-Herriot model, named after the authors of the landmark paper with more than 1200 citations to date (according to Google Scholar) by Fay and Herriot (1979). For a comprehensive review of small area estimation, the readers are referred to the book by Jiang (2007) and Rao and Molina (2015).

We may be interested in the high dimensional parameters (random effects) $\theta_i$ and/or the hyperparameters $\beta$ and $A$.  The estimation problem can be addressed using either Bayesian or linear mixed model classical approach.  When hyperparameters are known, both the Bayesian and linear mixed model classical approaches use conditional distribution of  $\theta_i$  given the data for point estimation and measuring uncertainty of the point estimator.  To elaborate, the posterior mean of $\theta_i$, the Bayesian point estimator, is identical to the best predictor of $\theta_i$.  Moreover, the posterior variance of $\theta_i$ is identical to the mean squared error of the best predictor.  When $A$ is known but $\beta$ is unknown, a flat prior is generally assumed for $\beta$ under the Bayesian approach. Interestingly, in this unknown $\beta$ case, the posterior mean and posterior variance of $\beta$  are identical to the maximum likelihood estimator of $\beta$ and the variance of the maximum likelihood estimator, respectively.  Moreover, the posterior mean and variance of $\theta_i$ are identical to the best linear unbiased predictor of $\theta_i$ and its mean squared error, respectively.

When both $\beta$ and $A$ are unknown, flat prior, i.e., $\pi (\beta,A)\propto 1,\;\beta\in \mathcal{R}^p, A>0$, is common though a few other priors for $A$ have been considered; see, e.g., Datta et al. (2005) and Morris and Tang (2011).  In a linear mixed model classical approach, different estimators of $A$ have been proposed and  the estimator of $\beta$ is obtained by plugging in an estimator of $A$ in the maximum likelihood estimator of $\beta$ when $A$ is known.  In this general case, the relationship between the Bayesian and linear mixed model classical approach is not clear.  The main goal of this paper is to understand the nature of such relationship.  In particular, we answer the following question:  For a given classical method of estimation of $A$, is it possible to find a prior on $A$ that will make the Bayesian solution closer to the classical solution in achieving multiple goals (i)-(v), described in Section 3, or a subset of these goals given in Theorem 2?

What would be the parameters of interest in setting the multiple goals? To this end, we first note that Morris and Tang (2011) pointed out the need for accurately estimating the shrinkage parameters $B_i=D_i/(A+D_i)$ as they appear linearly in the Bayes estimators of $\theta_i$, which are the prime parameters of interest in many applications like the small area estimation.  Moreover, the shrinkage parameters are good indicators of the strength of the prior on the random effects $\theta_i$.  Despite the importance of shrinkage parameters, relatively little research  has been conducted in order to understand the theoretical properties of existing estimators.  For the balanced case when $D_i=D,\; i=1,\cdots,m$, Morris (1983) proposed an exact unbiased estimator of $B=D/(A+D)$ and showed component-wise dominance of the resulting empirical Bayes estimator of $\theta_i$ under the joint distribution of $\{(y_i,\theta_i),\;i=1,\cdots,m\}$ when $p\le m-3.$ For the general unbalanced case, Hirose and Lahiri (2018) proposed an adjusted maximum likelihood  estimator of $B_i$ that satisfies multiple desirable properties. First, the method yields an estimator of $B_i$ that is strictly less than 1, which prevents the overshrinking problem in the related empirical best linear unbiased predictor or simply empirical best predictor of $\theta_i$.  Secondly, this adjusted maximum likelihood estimator of $B_i$ has the smallest bias among all existing rival estimators in the higher order asymptotic sense.  Thirdly, when this adjusted maximum likelihood method is used, second-order unbiased estimator of mean squared error of empirical best linear unbiased predictor can be produced in a straightforward way without additional bias corrections that are necessary for other existing variance component estimation methods. 
For prior work on the adjusted maximum likelihood method, the readers are referred to Lahiri and Li (2009), Li and Lahiri (2010), Yoshimori and Lahiri (2014a,b), Hirose and Lahiri (2018), and Hirose (2017,2019). 

As stated in Morris and Tang (2011), flat prior leads to admissible minimax estimators of the random effects for a special case of the model.  In Section 3, we show that the bias of the Bayes estimator of $B_i$, under the flat prior and the two-level model, is $O(m^{-1})$ except for the balanced case when it is of lower order $o(m^{-1})$. Thus, in general, the Bayes estimator of $B_i$, under the flat prior, has more bias than the adjusted maximum likelihood estimator of  Hirose and Lahiri (2018) in the higher order asymptotic sense.  In this section, we propose a prior for the hyperparameters that leads to the Bayes estimator of $B_i$  with bias of lower order $o(m^{-1})$ and thus is on par with the adjusted maximum likelihood of Hirose and Lahiri (2018).  Interestingly, this prior also makes the resulting Bayesian method much closer to the Hirose-Lahiri\rq{}s empirical best linear unbiased prediction method in multiple sense. 
In particular, the posterior variance of the random effect $\theta_i$, under the proposed prior, is identical to both the Taylor series and parametric bootstrap second-order mean squared error estimators of Hirose and Lahiri (2018) in the higher order asymptotic sense.  To our knowledge, we establish for the first time the relationship between the Bayesian posterior variance and parametric bootstrap mean squared error estimator in this higher-order asymptotic sense.

The outline of the paper is as follows.  In Section 2, we first introduce a classical method for the  two level model by proposing a general adjustment factor in estimating $A$.  We show how the method is related to the commonly used residual maximum likelihood method for a given choice of the adjustment factor. We then construct a prior, called a multi-goal prior, that provides a Bayesian solution close (with respect to several properties in higher order asymptotic sense) to classical solution in order to estimate the hyperparameters and random effects.  Section 3 discusses prior choice for an important special case considered by Hirose and Lahiri (2018).  In addition to the multiple properties discussed in Section 2, this section develops a unique multi-goal prior that establishes a relationship of the posterior variances of the random effects with the Hirose-Lahiri Taylor series and parametric bootstrap mean squared error estimators that do not require the usual complex bias corrections. We reiterate that this paper demonstrates for the first time how to bring the Bayesian and classical parametric bootstrap methods closer in the context of random effects models.  In Section 4, we compare the proposed  multi-goal prior with the superharmonic prior using a real life data.  In Section 5, we discuss issues in extending our results to a general model.  All the technical proofs are deferred to the Appendix.

\section{Prior Choice for reconciliation of the  Bayesian and classical approach}
In this section, we first introduce a general classical method for estimation of hyperparameters and random effects in the two-level Normal hierarchical model.  Then we construct prior for the hyperparameters so that the corresponding Bayesian method is identical to the classical method  in the higher order asymptotic sense with respect to multiple properties. 

We first introduce the empirical best linear unbiased predictor of $\theta_i$ when the variance component $A$ is estimated by a general adjusted maximum likelihood method.  To this end, we define mean squared error of a given predictor $\hat{\theta}_i$ of $\theta_i$ as $M_i(\hat{\theta}_i)=E(\hat{\theta}_i-\theta_i)^2$, where the expectation is with respect to the joint distribution of $y=(y_1,\cdots,y_m)^{\prime}$ and $\theta=(\theta_1,\cdots,\theta_m)^{\prime}$ under the two-level normal model. 
The best linear unbiased predictor $\hat{\theta}_i^{BLUP}$ of $\theta_i$, which minimizes $M_i(\hat{\theta}_i)$ among all linear unbiased predictors $\hat\theta_i$, is given by 
$\hat{\theta}_i^{BLUP} (A)=(1-B_i)y_i+B_i x^{\prime}_i\hat{\beta}(A),$ 
where $B_i\equiv B_i(A)=D_i/(A+D_i)$ is the shrinkage factor and $\hat{\beta}(A)=(X^{\prime}{V}^{-1}X)^{-1}X^{\prime}{V}^{-1}y$ is the weighted least square estimator of $\beta$ when $A$ is known. In this formula, $X^{\prime}=(x_1,\cdots,x_m)$ denotes $p\times m$ matrix of known auxiliary variables and $V=\mbox{diag}(A+D_1,\cdots,A+D_m)$ denotes a $m\times m$ diagonal covariance matrix of $y$. 

We consider the following general adjusted maximum likelihood estimator $\hat{A}_{i;G}$ of $A$ : 
\begin{align}
\hat{A}_{i;G}=\argmax_{0 \le A<\infty} h_{i;G}(A)L_{RE}(A), \label{ad.est}
\end{align}
where the general adjustment factor $h_{i;G}(A)$ satisfies Condition R5 in Appendix A. 
Note that maximum likelihood, residual maximum likelihood and different adjusted maximum likelihood estimators of $A$ can be produced using suitable choices of $h_{i;G}(A)$. Plugging in  $\hat{A}_{i;G}$ for $A$ in the best linear unbiased predictor, one obtains an empirical best linear unbiased predictor $\hat{\theta}_i^{EB}(\hat{A}_{i;G})$ of $\theta_i$. 

Since the residual maximum likelihood estimator of $A$ has the lowest bias among existing estimators in the higher-order asymptotic sense, it is of interest to establish a relationship between the general adjusted maximum likelihood estimator and the residual maximum likelihood estimator. We describe such relationship in  Theorem \ref{L1}; see Appendix A.1 for a proof.

\begin{theorem}
\label{L1}
Under regularity conditions R1-R5, 
$$\hat{A}_{i;G}-\hat{A}_{RE}=\frac{2\tilde l_{i;G}^{(1)}(A)}{tr[V^{-2}]}+o_p(m^{-1}),$$ 
\end{theorem}
where $\tilde l_{i;G}^{(1)}(A)=\frac{\partial \log h_{i;G}(A)}{\partial A}$.

We now present Theorem \ref{rel} for constructing a prior, starting from a given adjustment factor $h_{i,G}(A)$, in order to bring the resulting Bayesian method closer to the classical method with respect to three criteria.  To this end, let $p (\beta, A)$ denote the prior for $(\beta,A)$.  Following Datta et al. (2005), we assume 
$p (\beta, A)\propto \pi (A)$ and introduce the following notations to be used throughout the paper: 
\begin{align*} 
&\hat b_1=\frac{\partial B_i}{\partial A}\Big |_{\hat{A}_{RE}},\ \hat b_2=\frac{\partial^2 B_i}{\partial A^2}\Big |_{\hat{A}_{RE}},\ \hat \rho_1 =\frac{\partial \log \pi(A)}{\partial A}\Big |_{\hat{A}_{RE}},\\ 
&\hat h_2=-\frac{1}{m}\frac{\partial^2 l_{RE}}{\partial A^2}\Big |_{\hat{A}_{RE}}=\frac{tr[V^{-2}]}{2m}+o_p(m^{-1}),\\ 
&\hat h_3=-\frac{1}{m}\frac{\partial^3 l_{RE}}{\partial A^3}\Big |_{\hat{A}_{RE}}=-\frac{2tr[V^{-3}]}{m}+o_p(m^{-1}),
\end{align*}
where $\hat A_{RE}$ is the residual maximum likelihood estimator of $A$, and $l_{RE}$ is the logarithm of residual likelihood.
\begin{theorem}
\label{rel}
Under Regularity Conditions R1-R5, if $p(\beta,A)\propto \pi_{i;G}(A)$ and 
\begin{align}\pi_{i;G}(A)\propto (A+D_i)tr(V^{-2}){h}_{i;G}(A),\label{g.p}\end{align}
we have;
\begin{align*}
&(i) \hat{B}_i^{GHB}=\hat{B}_i(\hat{A}_{i;G})+o_{p}(m^{-1});\\
&(ii) \hat{V}_i^{GHB}=V[B_i|y]=Var( \hat B_i(\hat{A}_{i;G}))+o_p(m^{-1});\\
&(iii) \hat{\theta}_i^{GHB}=\hat{\theta}_i(\hat{A}_{i;G})+o_{p}(m^{-1}).
\end{align*}
\end{theorem}
The proof of Theorem \ref{rel} is deferred to Appendix A.2.

\begin{remark}
\label{pri.cond}
We have several remarks on the general multi-goal prior given by (\ref{g.p}).  
\begin{description}
\item[(a)] Theorem \ref{rel} is valid for multiple choices of $h_{i;G}$.

\item[(b)]
There exists at least one strictly positive estimate of $A$ if $h_{i;G}(A)>0$ and 
\begin{align}
h_{i;G}(A)=o(A^{(m-p)/2}),\label{F.ec}
\end{align}
for large $A$ under R6-7.

\item[(c)] Note that $h_{i;G}(A)$ may not qualify as a bonafide prior since it may result in an improper posterior;  see Yoshimori and Lahiri (2014b) for an example.  However, if we restrict the class of priors to $h_{i;G}(A)=(A+D_i)^s$ for some $s>0$, we show in Appendix B.1 that $h_{i;G}(A)=o(A^{(m-p-2)/2})$ is a sufficient condition for the propriety of posterior and hence can serve as a prior for $A$.

On the other hand, it is straightforward to show that $\pi_{i;G}(A)$ given by (\ref{g.p}) with $h_{i;G}(A)=o(A^{(m-p)/2})$ yields proper posterior because of multiplication of $h_{i;G}(A)$ by $(A+D_i)tr(V^{-2})$. 
In either case, Theorem \ref{rel} can facilitate users for selecting an adjusment factor in the emprical best linear unbiased prediction approach or prior in the Bayesian approach.

\end{description}
\end{remark}

\section{Multi-Goal Prior for an important special case}

Hirose and Lahiri (2018) put forward a classical approach for an important choice of $h_{i;G}(A)$ that satisfies the following desirable properties under regularity conditions R1-R7: 
 
\begin{description}
\item [1.] It is desirable to have a second-order unbiased estimator of $B_i$, i.e., $E(\hat B_i)=B_i+o(m^{-1})$.
\item [2.]$0<\mbox{inf}_{m\ge 1}\hat B_i\le \mbox{sup}_{m\ge 1}\hat B_i<1$ (a.s.) 
for protecting the empirical best linear unbiased predictor from over-shrinking to the regression estimator.
\item [3.] It is desirable to obtain a simple second-order unbiased Taylor series mean squared error estimator of the empirical best linear unbiased predictor without any bias correction; that is, $E[\hat{M}_{i}(\hat A_i)]=M_i(\hat{\theta}_i^{EB})+o(m^{-1}).$
\item [4.] It is desirable to produce a strictly positive second-order unbiased single parametric bootstrap mean squared error estimator without any bias-correction,
\end{description}
where $\hat{M}_i(\hat{A}_i)$ denotes a estimator of mean squared error of $\hat\theta_i^{EB}(\hat{A})$. 

Let $\hat{A}_{i;MG}$, $\hat{B}_{i;MG}$, $\hat{\theta}_{i;MG}^{EB}$, $\hat{M}_{i;MG}$, $\hat{M}_{i;MG}^{boot}$ be the Hirose--Lahiri\rq{}s estimators of $A, B_i,$ the empirical best linear unbiased predictor of $ \theta_i$, Taylor series and parametric bootstrap estimators of the mean squared error of the empirical best linear unbiased predictor, respectively. They are given by
\begin{align*} 
\hat {A}_{i;MG}=\argmax_{0< A <\infty}& \tilde h_i(A)L_{RE}(A),\\ 
\hat {B}_{i;MG}=\hat{B}_i(\hat{A}_{i;MG}),& \  \hat {\theta}_{i;MG}^{EB}=\hat{\theta}_i^{EB}(\hat{A}_{i;MG}), \\
\hat{M}_{i;MG}=\hat{M}_i(\hat{A}_{i;MG}),& \ \hat{M}_{i;MG}^{boot}=E_*[\{\hat{\theta}_i(\hat{A}_{i;MG}^*,y^{*})-\theta_i^*)\}^2],
\end{align*}
where $\tilde h_i(A)=h_{+}(A)(A+D_i)$ with $m>p+2$; $h_{+}(A)$ satisfies Conditions R6-R7 in Appendix A; 
$\theta_i^{*}=x_i^{\prime}\hat \beta(\hat{A}_{1;MG},\ldots,\hat{A}_{m;MG})+u_i^*$ with $u_i^* \sim^{ind.} N(0,\hat{A}_{i;MG})$; $E_*$  is expectation with respect to the two-level Normal hierarchical model with $\beta$ and $A$ replaced by $\hat\beta(\hat{A}_{1;MG},\ldots,\hat{A}_{m;MG})$ and $\hat A_{i;MG}$, respectively. Note that the choice of $h_{+}(A)$ is not unique in general.  One can use the choice given in Yoshimori and Lahiri (2014a). 

The following corollary follows from Theorem \ref{L1}, Hirose and Lahiri (2018) and the fact that $\frac{\partial \hat \beta(A)}{\partial A}=O_p(m^{-1/2})$. 
\begin{corollary}
\label{cor}
Using the regularity conditions,
\begin{align*}
&(i) \hat{A}_{i;MG}-\hat{A}_{RE}=O_{p}(m^{-1});\\
&(ii) x_i^{\prime}\hat{\beta}(\hat{A}_{1;MG},\ldots,\hat{A}_{m;MG})-x_i^{\prime}\hat{\beta}(\hat{A}_{RE})=o_{p}(m^{-1}). 
\end{align*}
\end{corollary}

In this section, we suggest a Bayesian approach that is close to the classical approach to achieve multiple goals in the higher-order asymptotic sense. To this end, we seek a multi-goal prior on the hyperparameters $(\beta,A)$ that satisfies all the following properties simultaneously:
\begin{description}
\item[(i)] $\hat{B}_i^{HB}\equiv E[B_i|y]=\hat{B}_{i,MG}+o_p(m^{-1})$;
\item[(ii)] $V[B_i|y]=Var( \hat B_{i;MG} )+o_p(m^{-1})$;
\item[(iii)] $\hat{\theta}_i^{HB}\equiv E[\theta_i|y]=\hat{\theta}_{i,MG}+o_p(m^{-1})$;
\item [(iv)] $V[\theta_i|y]=\hat{M}_{i;MG}+o_p(m^{-1})$;
\item [(v)] $V[\theta_i|y]=\hat{M}_{i;MG}^{boot}+o_p(m^{-1})$.
\end{description}

First we prepare the following result, which follows from Corollary \ref{cor} (i) and Hirose and Lahiri (2018): 
\begin{align}
\hat{B}_i(\hat{A}_{i;MG})-\hat{B}_i(\hat{A}_{RE})&=(\hat{A}_{i,MG}-\hat{A}_{RE})\hat b_1+o_p(m^{-1})\notag\\
&=\{E[\hat{A}_{i;MG}-A]-E[\hat{A}_{RE}-A]\} b_1+o_p(m^{-1})\notag\\
&=-\frac{2D_i}{tr[V^{-2}](A+D_i)^3}+o_p(m^{-1}).\label{bias.B}
\end{align}

If we use the flat prior $\pi(A)\propto 1$, we get the following result using equation (21) of Datta et al. (2005) with $b(A)=B_i(A)$ and equation (\ref{bias.B}): 
$$E[B_i|y]=\hat B_i(\hat A_{MG})+\frac{4D_i}{tr[V^2](A+D_i)^2}\left[\frac{1}{A+D_i}-\frac{tr[V^{-3}]}{tr[V^{-2}]}\right]+o_p(m^{-1}).$$ 
This result emphasizes that the flat prior $\pi(A)\propto 1$ cannot achieve Property (i) except for balanced case ($D_i=D$ for all $i$). We, therefore, seek a prior $\pi (A)$ to satisfy Property (i), even in unbalanced case. 
To this end, we also use the following result (\ref{B.HB}) given in (21) of Datta et al. (2005) with $b(A)=B_i(A)$:
\begin{align}
E[B_i|y]=\hat{B}_i(\hat{A}_{RE})+\frac{1}{2m\hat{h}_2}\left(\hat b_2-\frac{\hat{h}_3}{\hat{h}_2}\hat b_1 \right)+\frac{\hat{b}_1}{m\hat{h}_2}\hat \rho_1+o_p(m^{-1}).\label{B.HB}
\end{align}
It is evident from equations (\ref{bias.B}) and (\ref{B.HB}) that our desired prior must satisfy the following differential equation, up to the order of $O(m^{-1})$:
\begin{align}
\frac{1}{2m{h}_2}\left( b_2-\frac{{h}_3}{{h}_2} b_1 \right)+\frac{{b}_1}{m{h}_2} \rho_1=-\frac{2D_i}{tr[V^{-2}](A+D_i)^3}.\label{diff1}
\end{align}

Note that the differential equation (\ref{diff1}) is equivalent to the following differential equation, up to the order of $O_p(m^{-1})$;
\begin{align}
\rho_1=\frac{\partial \log \pi(A)}{\partial A}&=-\frac{m  h_2}{ b_1}\frac{2D}{tr[V^{-2}](A+D_i)^3}-\frac{1}{2}\left[\frac{ b_2}{ b_1}-\frac{ h_3}{ h_2} \right]\notag\\
&=\frac{2}{A+D_i}-\frac{2tr[V^{-3}]}{tr[V^{-2}]}.\label{difffinal}
\end{align}

Hence, we obtain a solution to differential equation (\ref{difffinal}) as follows: 
\begin{eqnarray}
\pi(A)\propto (A+D_i)^2tr[V^{-2}]. \label{MG.pri0}
\end{eqnarray}

Note that the prior (\ref{MG.pri0}) depends on $i$. Therefore, we redefine it as:
\begin{eqnarray}
\pi_i(A)\propto (A+D_i)^2tr[V^{-2}]. \label{MG.pri}
\end{eqnarray}

\begin{remark}
\label{R1}
We have several important remarks on the prior (\ref{MG.pri}). 
\begin{description}
\item[(a)] The prior satisfies the rest of Properties (ii)-(v) simultaneously, as shown in Appendix B.2. 
It is remarkable that $\pi_i(A)$ given by (\ref{MG.pri}) is the unique prior to achive Properties (i)-(v) simultaneously, up to the order of $O_p(m^{-1})$, since $E[g_{1i}(A)|y]=g_{1i}(\hat A_{i;MG})+o_p(m^{-1})$ shown in (\ref{g1.exp}).

\item[(b)] The prior given by equation (\ref{MG.pri}) reduces to the Stein's super-harmonic prior for the balanced case $D_i=D,\;i=1,\cdots,m$, up to the order of $O_p(m^{-1})$.

\item[(c)] Datta et al. (2005) found the same prior by matching (in a higher order asymptotic sense) expected value of the posterior variance of $\theta_i$ with the mean squared error of the empirical best linear unbiased predictor with the residual maximum likelihood estimator used for the variance component $A$. It is interesting to note that the same prior achieves multiple goals, a fact gone unnoticed.

\item [(d)] From the result of Ganesh and Lahiri (2008), the prior 
$$\pi (A)\propto \frac{\sum \{1/(A+D_i)^2\}}{\sum \omega_i\{D_i^2/(A+D_i)^2\}}$$ also satisfies  
$\sum_{i}^m \omega_i E[{V}(\theta_i|y)-MSE[\hat \theta_i(\hat{A}_{i;MG})]]=o(m^{-1}).$
\end{description}
\end{remark}

\section{Data Analysis}
In this section, using the 1993 Small Area Income and Poverty Estimates (SAIPE) data set, we demonstrate that our proposed multi-goal prior(MGP) performs better than the superharmonic prior (SHP) in producing Bayesian solutions closer to the multi-goal classical solutions of Hirose and Lahiri (2018).  The SAIPE data we use here is from Bell and Franco (2017), available at \url{https://www.census.gov/srd/csrmreports/byyear.html}. The data contains direct poverty rates($y_i$), associated sampling variances ($D_i$), and auxiliary variables ($x_i$) derived from administrative and census data for the 50 states and the District of Columbia.  Much has been written about SAIPE over the years. See, for instance, the recent book chapter by Bell et al. (2016). 

First consider the estimation of the shrinkage parameters $B_i$ for all the states. Fig 1 displays classical multi-goal estimates $\hat B_{i;MG}$ and Bayes estimates of $B_i$ under the superharmonic  and the multi-goal priors for all the states arranged in decreasing order of $\hat B_{i;MG}$. Note that the Bayes estimate of $B_i$ is an one-dimensional integral, which is approximated by numerical integration using the R function \lq\lq{}adaptIntegrate\rq\rq{}. Overall, the Bayes estimates under the multi-goal prior are closer to the classical estimates (MGF) than the superharmonic prior.

Next, in Fig 2, we compare the mean squared error estimates by Taylor series (MGF) and parametric bootstrap (PB MG) of Hirose and Lahiri (2018) with the posterior variances under the two different priors.  The parametric bootstrap mean squared error estimates use $10^4$ bootstrap samples. The two mean squared error estimates are virtually identical.  Again our posterior variances under the  multi-goal prior are much closer to the mean squared error estimates than the corresponding posterior variances under the superharmonic prior.

\begin{figure}[ht]
\includegraphics[bb=0 0 1920 1082,scale=0.2,clip]{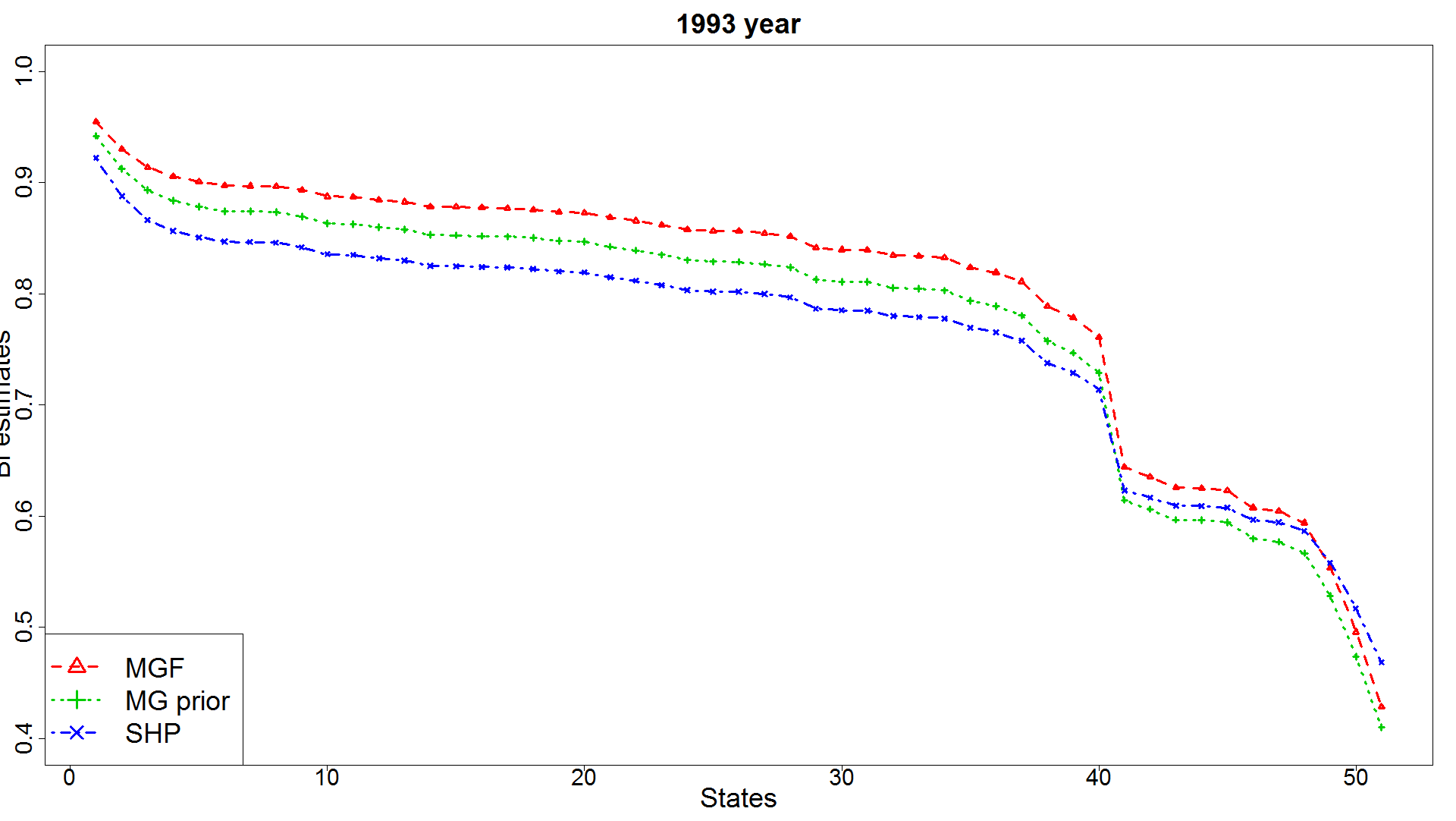}
\caption{$B_i$ estimates (MGF:$\hat B_{i;MG}$, MGP: $E_{MG}[B_i|y]$, SHP:$E_{SHP}[B_i|y]$)}
\label{Bi.93}
\end{figure}

\begin{figure}[ht]
\includegraphics[bb=0 0 1920 1082,scale=0.2,clip]{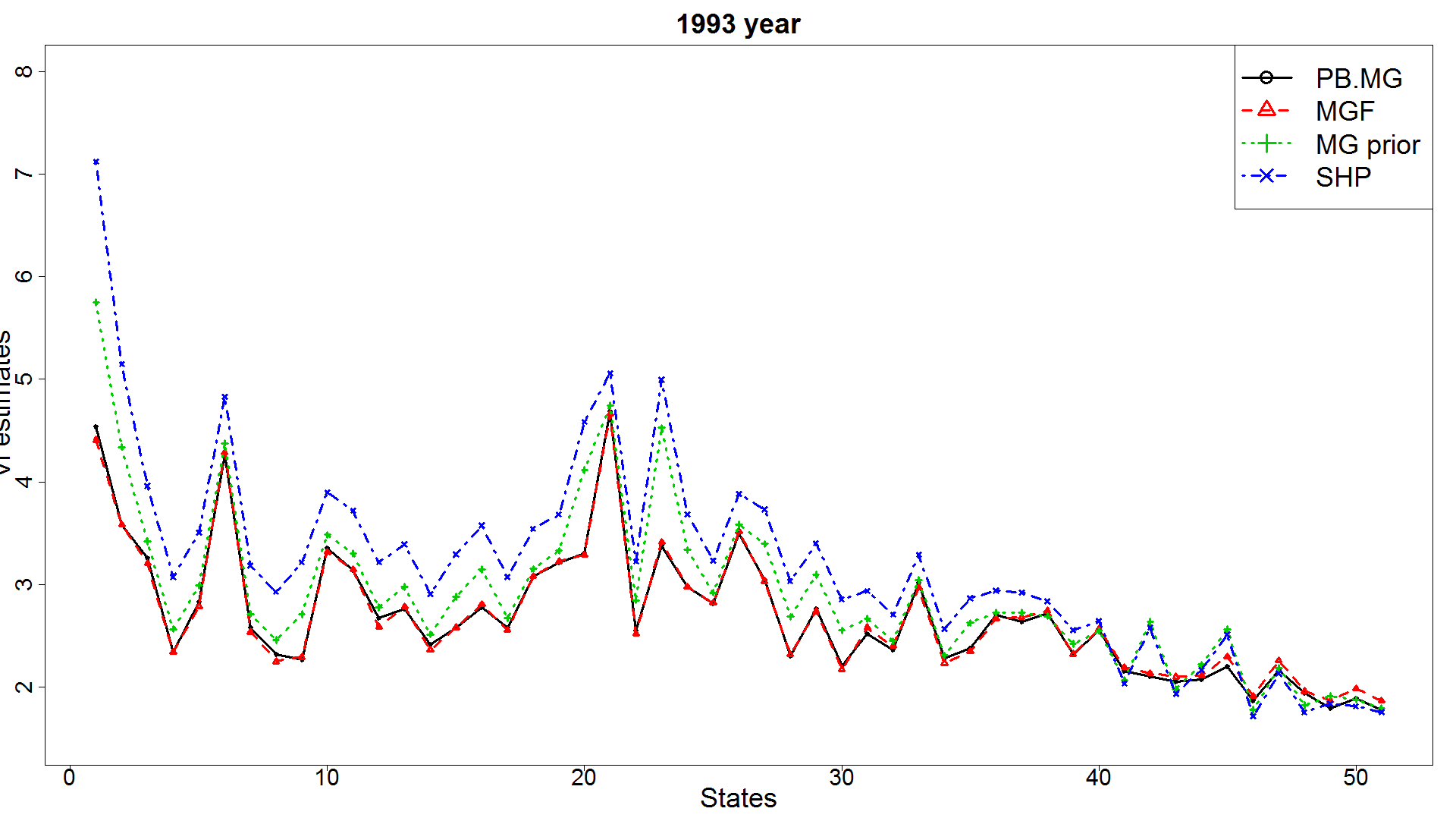}
\caption{MSE estimates 
(PB.MG:$\hat M^*_{i;MG}$, MGF:$\hat M_{i;MG}$, MG Prior:$V_{MG}[\theta_i|y]$, SHP:$V_{SHP}[\theta_i|y]$)
}
\label{Vi93}
\end{figure}

\section{Discussion}  

Can we extend our results to a general linear mixed model? To answer this question, we consider the following nested error regression model considered by Battese et al. (1988): 
\begin{eqnarray}
y_{ij}=\theta_{ij}+e_{ij}=x_{ij}^{\prime}\beta+v_i+e_{ij}, \ (i=1,\ldots,m;\ j=1,\ldots,n_i),\label{NERM}
\end{eqnarray}
where $\{v_1\ldots,v_m\}$ and $\{e_1,\ldots, e_m\}$ are independent with $v_i{\sim}N(0,\sigma_v^2)$ and $e_i{\sim}N(0,\sigma_e^2)$; $x_{ij}$ is a $p$-dimensional vector of known auxiliary variables; $\beta\in \mathcal{R}^p$ is a $p$-dimensional vector of unknown regression coefficients; $\psi=(\sigma_v^2, \sigma_e^2)^{\prime}$ is an unknown variance component vector; $n_i$ is the number of observed unit level data in $i$-th area.

The condition for achieving desired property 1 given in Section 3, we need to solve the following system of differential equations with shrinkage factor $B_i=\sigma_e^2/(n_i\sigma_v^2+\sigma_e^2)$, under certain regularity conditions: 
\begin{align}
\left[\frac{\partial \log h_{i;G}(\psi)}{\partial \psi}\right]^{\prime}I_F^{-1}\left[\frac{\partial B_{i}(\psi)}{\partial \psi}\right]=&H(\psi),
\end{align}
where $$\frac{\partial \log h_{i;G}(\psi)}{\partial \psi}=\left(\frac{\partial \log h_{i;G}(\psi)}{\partial \sigma_v^2}, \frac{\partial \log h_{i;G}(\psi)}{\partial \sigma_e^2}\right)^{\prime},$$ 
$$H(\psi)=-\frac{1}{2}tr\left[\frac{\partial^2 B_{i}(\psi)}{\partial \psi^2}I_F^{-1}\right], \  
\frac{\partial B_{i}(\psi)}{\partial \psi}=\frac{n_i}{(n_i\sigma_v^2+\sigma_e^2)^2}(-\sigma_e^2, \sigma_v^2)^{\prime},$$ 
$$I_F^{-1}=\frac{2}{a}\left (
\begin{array}{cc}
\sum[(n_i-1)/\sigma_e^4+(n_i\sigma_v^2+\sigma_e^2)^{-2}] & -\sum n_i/(n_i\sigma_v^2+\sigma_e^2)^2\\
-\sum n_i/(n_i\sigma_v^2+\sigma_e^2)^2&\sum n_i^2/(n_i\sigma_v^2+\sigma_e^2)^2\\
\end{array}\right),$$ 
$$a=[\sum n_i^2/(n_i\sigma_v^2+\sigma_e^2)^2][\sum \{(n_i-1)/\sigma_e^4+(n_i\sigma_v^2+\sigma_e^2)^{-2}\}]-[\sum n_i/(n_i\sigma_v^2+\sigma_e^2)^2]^2.$$

If we use the following adjustment factor $h_{i;G}(\psi)$ for achieving desired property 1:
\begin{align}
\frac{\partial \log h_{i;G}(\psi)}{\partial \psi}=v{k},
\end{align}
for a given two dimensional fixed vector ${k}$, 
the solution of $v$ can be obtained as $$v=\frac{H(\psi)}{{k}^{\prime}I_F^{-1}\frac{\partial B_{i}(\psi)}{\partial {\psi}}}.$$ 
This solution thus leads to an appropriate adjustment factor satisfying $$\frac{\partial \log h_{i;G}(\psi)}{\partial \psi}=\frac{H(\psi)}{{k}^{\prime}I_F^{-1}\frac{\partial B_{i}(\psi)}{\partial {\psi}}}{k}.$$ 
Thus, there exist multiple solutions for $h_{i;G}(\psi)$ satisfying desired property 1 under the nested error regression model (\ref{NERM}). 
Further research is needed to identify a reasonable adjustment factor for the general linear mixed model and to establish a connection with the corresponding Bayesian approach.

\section*{Acknowledgements}
The first and second authors\rq{} research was supported by JSPS KAKENHI Grant Number 18K12758 and U.S. National Science Foundation Grant SES-1534413, respectively.

\appendix
\section{Appendix}
We assume the regularity conditions throughout this paper as follows: 

{\bf Regularity Conditions}
\begin{description}

\item {\it R1:} $\mbox{rank}(X)=p$ is bounded for large $m$;
\item {\it R2:} The elements of $X$ are uniformly bounded implying $\sup_{j\geq 1}x_j(X^{\prime}X)^{-1}x_j=O(m^{-1})$;
\item {\it R3:} $0<\inf_{i\geq 1}D_i\leq \sup_{i\geq 1}D_i<\infty$, $A\in (0,\infty)$;
\item {\it R4:} $|\hat{A}_{i}|<C_{ad}m^{\lambda}$, where $\hat{A}_{i}$ is an estimator of $A$ and $C_{ad}$ a generic positive constant and  $\lambda$ is  small positive constant. 
\end{description}

We also restrict the class of adjustment factors $h_{+}(A)$ and $h_{i;G}(A)$ that satisfy the following regularity conditions, as in Hirose and Lahiri (2018):
\begin{description}
\item {\it R5:}  $\log h_{i;G}(A)$ is free of $y$ and  four times continuously differentiable with respect to $A$.  Moreover, $\frac{\partial^k \log h_{i;G}(A)}{\partial A^k}$ is of order $O(1)$, respectively, for large $m$ with $k=0,1,2,3$;
\item {\it R6:}  $\log h_{+}(A)$ is free of $y$ and  four times continuously differentiable with respect to $A$.  Moreover, $\frac{\partial^k \log h_{+}(A)} {\partial A^k}$ is of order $o(1)$, for large $m$ with $k=0,1,2,3$;
\item {\it R7;}  $h_{+}(A)$ is a strictly positive on $A>0$ satisfying that $h_{+}(A)\Big|_{A=0}=0$ and $h_{+}(A)<C$ on $A>0$ with a generic positive constant $C$.

\end{description}

\subsection{Proof of Theorem \ref{L1}}

The result follows from an argument similar to the ones given in Das et al. (2004). We note that for the general adjusted maximum likelihood method (\ref{ad.est}), 
\begin{align}
l_{i;G}^{(1)}(\hat{A}_{i;G})-l_{i;G}^{(1)}({A})=&(\hat{A}_{i;G}-{A})E[l_{i;G}^{(2)}(A)]
+(\hat{A}_{i;G}-{A})\{l_{i;G}^{(2)}({A})-E[l_{i;G}^{(2)}({A})]\}\notag\\
&+\frac{1}{2}(\hat{A}_{i;G}-{A})^2l_{i;G}^{(3)}({A}_i^*),\label{pro1}
\end{align}
where $l_{i;G}^{(k)}(A)=\frac{\partial^k [\tilde{l}_{i;G}(A)+{l}_{RE}(A)]}{\partial A^k}$ for $k=1,2,3$ with $\tilde{l}_{i;G}(A)=\log h_{i;G}(A)$ and $\tilde{l}_{RE}(A)=\log L_{RE}(A)$. In addition, ${A}_i^*$ lies between $A$ and $\hat{A}_{i;G}$. 

Under regularity conditions, using results of Hirose and Lahiri (2018) and $l_{i;G}^{(1)}(\hat{A}_{i;G})=0$, we have
$\hat{A}_{i;G}-A=O_p(m^{-1/2})$, $\hat{A}_{i}^*-A=O_p(m^{-1/2})$, 
$l_{RE}^{(1)}(\hat{A}_{i;G})=-\tilde{l}_{i;G}^{(1)}(\hat{A}_{i;G})$, 
$E[l_{i;G}^{(2)}(A)]=E[l_{RE}^{(2)}(A)]+O(1)=-\frac{tr[V^{-2}]}{2}+O(1)$, 
$|l_{RE}^{(2)}({A})|=O_{p}(m)$, $|l_{RE}^{(3)}(A)|=O_{p}(m)$. 

Hence, (\ref{pro1}) yields:
\begin{align}
\hat{A}_{i;G}-\hat{A}_{RE}&=\hat{A}_{i;G}-A-(\hat{A}_{RE}-A)\notag\\
&=\frac{2}{tr[V^{-2}]}\tilde{l}_{i;G}^{(1)}
+\left\{\frac{2}{tr[V^{-2}]}\right\}^2\tilde{l}_{i;G}^{(1)}(A)\{l_{RE}^{(2)}({A})-E[l_{RE}^{(2)}({A})]\}\notag\\
&+\frac{1}{2}\left\{\frac{2}{tr[V^{-2}]}\right\}^3\{\tilde{l}_{i;G}^{(1)}(A)(\tilde{l}_{i;G}^{(1)}(A)+2{l}_{RE}^{(1)}(A))\}\{{l}_{i;G}^{(3)}(A)+o_p(m)\}.\label{pro2}
\end{align}

Using the fact that $l_{RE}^{(1)}({A})=o_p(m)$, 
$$(\ref{pro2})=\frac{2}{tr[V^{-2}]}\tilde{l}_{i;G}^{(1)}+o_p(m^{-1}).$$
Theorem \ref{L1} thus follows.

\subsection{Proof of Theorem \ref{rel}}

\begin{proof}{of part (i):}

Using Theorem \ref{L1}, we have 
\begin{align}
\hat B_i(\hat A_{i;G})&=\hat B_i(\hat A_{RE})-\tilde{l}_{i;G}^{(1)}(A)\frac{2B_i^2}{tr[V^{-2}]D_i}+o_p(m^{-1}).\label{B.G}
\end{align}

Hence, using (\ref{B.HB}) given in (21) of Datta et al. (2005), equation (2) implies that the following condition is required in order to satisfy $\hat{B}_{i}^{HB}=\hat{B}_i(\hat{A}_{i;G})$:
\begin{align}
\frac{1}{2m\hat{h}_2}\left(\hat b_2-\frac{\hat{h}_3}{\hat{h}_2}\hat b_1 \right)+\frac{\hat{b}_1}{m\hat{h}_2}\hat \rho_1=-\tilde{l}_{i;G}^{(1)}(A)\frac{2B_i^2}{tr[V^{-2}]D_i}.\label{sum1}
\end{align}

Equation (\ref{sum1}) reduces to:
\begin{align}
\frac{\partial \log \pi_{i;G}(A)}{\partial A}=\tilde{l}_{i;G}^{(1)}(A)+\frac{1}{A+D_i}-\frac{2tr[V^{-3}]}{tr[V^{-2}]}+o_p(m^{-1}).\label{diff2}
\end{align}

After solving the above differential equation, up to the order of $O_p(m^{-1})$, we obtain: 
$\pi_{i;G}(A)\propto h_{i;G}(A)(A+D_i)tr[V^{-2}].$ 
 
Part (i) follows from this result. 
\end{proof}

\begin{proof}{of part (ii):}
Under regularity conditions, Hirose and Lahiri (2018) proved the following result:
$$Var(\hat{B}_{i}(\hat A_{i;G}))=\frac{2D_i^2}{m tr[V^{-2}](A+D_i)^4}+o(m^{-1}).$$

Hence,  using the result of Datta et al. (2005),
\begin{align}
V(B_i|y)=&\frac{\hat b_1^2}{m\hat h_1}+o_p(m^{-1})\notag\\
=& \frac{2D_i^2}{m tr[V^{-2}](A+D_i)^4}+o_p(m^{-1})\notag\\
=&Var(\hat{B}_{i}(\hat A_{i;G}))+o_p(m^{-1}).\label{th2.p2}
\end{align}
Thus, the prior (\ref{g.p}) satisfies property (ii) from (\ref{th2.p2}).

\end{proof}

\begin{proof}{of Part (iii):}

Datta et al. (2005) obtain the following result: 
\begin{align}
&E[g_{1 i}(A)|y]=g_{1i}(\hat A_{RE})+g_{1\pi i}(\hat A_{RE})+o_p(m^{-1});\notag\\
&\theta_i^{HB}=y_i-\hat B_i (\hat A_{RE})\{y_i-x_i^{\prime}\hat \beta(\hat{A}_{RE})\}+
\frac{g_{1\pi i}(\hat{A}_{RE})}{D_i}\{y_i-x_i^{\prime}\hat \beta(\hat{A}_{RE})\}+o_p(m^{-1}),\label{HB}
\end{align}
where \begin{align}
g_{1\pi i}(\hat A_{RE})=\frac{B_i^2}{m\hat h_2}\left(\hat \rho_1-\frac{1}{\hat A_{RE}+D_i}-\frac{\hat h_3}{2\hat h_2}  \right).\label{gpi}
\end{align}

Using (\ref{sum1}), we obtain
\begin{align}
g_{1\pi}(\hat{A}_{RE})=&\frac{B_i^2}{m\hat h_2}\tilde{l}_{i;G}^{(1)}(A)+o_p(m^{-1})\notag\\
=&\frac{2B_i^2}{tr[V^{-2}]}\tilde{l}_{i;G}^{(1)}+o_p(m^{-1}).\label{g1pi.G}
\end{align}

Hence, using Theorem \ref{L1}, (\ref{B.G}), (\ref{HB}),  (\ref{g1pi.G}) and the fact that $\partial \hat \beta(A)/\partial A=O_p(m^{-1/2})$, we have, for large $m$,
\begin{align*}
\theta_i^{GHB}=&y_i-\hat B_i (\hat A_{i;G})\{y_i-x_i^{\prime}\hat \beta(\hat{A}_{i;G})\}
+\{\hat B_i (\hat A_{i;G})-\hat B_i (\hat A_{RE})\}\{y_i-x_i^{\prime}\hat \beta(\hat{A}_{i;G})\}\\
&+\frac{2B_i^2}{tr[V^{-2}]D_i}\tilde{l}_{i;G}^{(1)}\{y_i-x_i^{\prime}\hat \beta(\hat{A}_{i;G})\}+o_p(m^{-1})\\
=&y_i-\hat B_i (\hat A_{i;G})\{y_i-x_i^{\prime}\hat \beta(\hat{A}_{i;G})\}+o_p(m^{-1}).
\end{align*}

This completes the proof of part (iii).
\end{proof}

\section{Appendix}

\subsection{Proof of Remark \ref{pri.cond} (c)}
We show that if we use $h_{i;G}(A)$ alone as a prior, $h_{i;G}(A)=o(A^{(m-p-2)/2})$ is a sufficient condition for the propriety of posterior in a constrained class of adjustment factors $h_{i:G}(A)=(A+D_i)^s$ for some $s>0$ and fixed $m$.  
We note that 
\begin{align}
\int_{0}^{\infty}L_{RE}(A)h_{i;G}(A)dA&\leq C\int_{0}^{\infty}(A+\inf_i D_i)^{-m/2}(A+\sup_i D_i)^{p/2+s}dA\notag\\
&=C\int_{0}^{\infty}\left[\frac{(A+\sup_i D_i)}{(A+\inf_i D_i)}\right]^{m/2}(A+\sup_i D_i)^{-m/2+p/2+s}dA\notag\\
&\leq C\int_{\sup_i D_i}^{\infty}t^{-m/2+p/2+s}dt\label{impro}.
\end{align}

It is evident that the condition $s<{(m-p-2)}/{2}$ achieves $(\ref{impro})<\infty$. 
Thus, the condition $h_{i;G}(A)=o(A^{(m-p-2)/2})$ is a sufficient condition for it to be a bonafide prior for large $A$.

The following inequality shows that $\pi_{i;G}(A)$ could be a prior if the condition $h_{i;G}(A)=o(A^{(m-p)/2})$ is met. 
\begin{align}
\int_{0}^{\infty}L_{RE}(A)\pi_{i;G}(A)dA&\leq C\int_{0}^{\infty}(A+\inf_i D_i)^{-m/2-2}(A+\sup_i D_i)^{p/2+1+s}dA\notag\\
&=C\int_{0}^{\infty}\left[\frac{(A+\sup_i D_i)}{(A+\inf_i D_i)}\right]^{m/2+2}(A+\sup_i D_i)^{-m/2-2+p/2+1+s}dA\notag\\
&\leq C\int_{\sup_i D_i}^{\infty}t^{-m/2+p/2-1+s}dt\label{impro2}.
\end{align}

Hence, if $h_{i;G}(A)$ in $\pi_{i;G}(A)$ satisfies $s<{(m-p)}/{2}$, then we have $(\ref{impro2})<\infty$. 
Thus, the condition $h_{i;G}(A)=o(A^{(m-p)/2})$ is a sufficient condition for $\pi_{i;G}(A)$ being a bonafide prior in a Bayesian method, as well as an adjustment factor in an adjusted maximum likelihood method.

\subsection{Proof of Remark \ref{R1} (a)}
We show that the prior (\ref{MG.pri}) achieves (ii)-(v). 

\begin{proof}{of (ii):}

From the result of Datta et al. (2005) and Hirose and Lahiri (2018), 
\begin{align}
V(B_i|y)=&\frac{\hat b_1^2}{m\hat h_1}+o_p(m^{-1})\notag\\
=& \frac{2D_i^2}{m tr[V^{-2}](A+D_i)^4}+o_p(m^{-1})\notag\\
=&Var(\hat{B}_{i;MG})+o_p(m^{-1}).\label{th1.p2}
\end{align}
Hence, the prior achieve the property (ii) from (\ref{th1.p2}).
\end{proof}

\begin{proof}{of (iii):}

Using (\ref{bias.B}), it is straightforward to show: 
$$g_{1i}(\hat A_{i;MG})-g_{1i}(\hat A_{RE})=\frac{2D_i^2}{tr[V^{-2}](A+D_i)^3}+o_p(m^{-1}).$$

Using (\ref{diff1}) and (\ref{gpi}), we obtain the following after some algebra:
\begin{align}
\label{g1pi}
g_{1i}(\hat A_{i;MG})=g_{1i}(\hat A_{RE})+g_{1\pi i}(\hat{A}_{RE})+o_p(m^{-1}).
\end{align}

Using (\ref{HB}), Corollary \ref{L1} (ii) and (\ref{g1pi}), we get:
\begin{align}
\theta_i^{HB}=&y_i-\hat B_i (\hat A_{i;MG})\{y_i-x_i^{\prime}\hat \beta(\hat{A}_{i;MG})\}\notag\\
&+\{\hat B_i (\hat A_{i;MG})-\hat B_i (\hat A_{RE})\}\{y_i-x_i^{\prime}\hat \beta(\hat{A}_{i;MG})\}\notag\\
&+\{\hat B_i (\hat A_{RE})-\hat B_i (\hat A_{i;MG})\}\{y_i-x_i^{\prime}\hat \beta(\hat{A}_{i;MG})\}+o_p(m^{-1})\notag\\
=&\hat{\theta}_{i;MG}^{EB}+o_p(m^{-1}).\label{th1.p3}
\end{align}

Property (iii) thus follows from the result (\ref{th1.p3}). 
\end{proof}

\begin{proof}{of (iv)-(v):}

Using (\ref{g1pi}), we get
\begin{align}
E[g_{1i}(A)|y]=g_{1i}(\hat A_{i;MG})+o_p(m^{-1}).\label{g1.exp} 
\end{align}

Datta et al. (2005) obtained the following results:
\begin{align}
V[\theta_i|y]=&g_{1i}(\hat A_{RE})+g_{1\pi i}(\hat{A}_{RE})+g_{2i}(\hat A_{RE})+g_{4i}(\hat A_{RE};y_i)+o_p(m^{-1}).\label{VHB}
\end{align}

Using the result given in Butar and Lahiri (2003), Hirose and Lahiri (2018), (\ref{g1pi}) and (\ref{g1.exp}), we get
\begin{align}
V[\theta_i|y]=&g_{1i}(\hat A_{i;MG})+g_{2i}(\hat A_{i;MG})+g_{3i}(\hat A_{i;MG})+o_p(m^{-1})\notag\\
=&\hat M_i(\hat{A}_{i;MG})+o_p(m^{-1})\notag\\
=&{M}_{i}(\hat \theta_{i;MG}^{EB})+o_p(m^{-1})\notag\\
=&\hat{M}_{i;MG}^{boot}+o_p(m^{-1}).\label{th1.p45} 
\end{align}
Equation (\ref{th1.p45}) implies that the prior (\ref{MG.pri}) also satisfies (iv)-(v) simultaneously. 
\end{proof}

\end{document}